# Rotorcraft RPM on Mars
Philip Blanco, Grossmont College, El Cajon, CA 92020.

The *Ingenuity* helicopter test flights on Mars in April 2021 marked the first time a powered aircraft has flown on another world. Students who have access to model helicopters and drones may wonder, could those hover on the Red Planet? The answer can be found in this journal, using appropriate scaling of the surface gravity and atmospheric density to Martian values. Liebl[1] provided physical arguments to model a toy helicopter's lifting thrust force $F_T$ as a function of air density $\rho_{atm}$, rotor radius $R$, and the rotors' cyclic frequency $f$ as

$$F_T = kR^4 \rho_{atm} f^2. \qquad (1)$$

This relation can also be derived from dimensional analysis. The dimensionless constant $k$ depends on the number, pitch, and shape of the rotor blades, which we assume do not change between terrestrial and Martian environments. We can therefore determine $k$ in advance by testing *Ingenuity* on Earth.

What rotor frequency is needed for hovering flight, when $F_T$ balances the helicopter's weight $mg$? Then,

$$F_T = kR^4 \rho_{atm} f^2 = mg \Rightarrow f = \sqrt{\frac{m}{kR^4}} \sqrt{\frac{g}{\rho_{atm}}} . \qquad (2)$$

The last square-root factor depends only on planetary properties. On the surface of Mars, $g$ = 3.7 m/s² = 0.38 × Earth's, and $\rho_{atm}$ = 0.020 kg/m³ = 0.017 × Earth's.[2] Therefore, from Eq. (2) the required rotor frequency $f$ for hovering flight on Mars should be $\sqrt{0.38/0.017}$ = 4.7 times that on Earth. This accords with a report that *Ingenuity*'s rotors must spin at ≈ 2400 rpm on Mars, compared to ≈ 500 rpm on Earth.[3]

Furthermore, with $m$ = 1.8 kg and $R$ = 0.60 m for *Ingenuity*,[4] we find $k$ ≈ 1.6 from Eq. (2). Students can then use Eq. (2) and Ref. 2 to calculate the rotor frequency for an *Ingenuity*-like helicopter to hover above the surface of Titan, or on other worlds. For comparison, they can also measure $k$ for a toy helicopter/drone as a laboratory project.[1] I hope that this exercise shows how performance values for sophisticated vehicles can often be estimated by simple scaling, as opposed to being pulled out of ... *thin air*!